\documentclass[a4paper]{panl}
\usepackage{cite}
\usepackage{wrapfig}
\usepackage{graphicx}
\usepackage{amssymb}
\usepackage{amsfonts}
\usepackage{amsmath}
\usepackage{longtable}
\usepackage{rotating}
\usepackage{lscape}
\usepackage{epsfig}
\usepackage{multirow}
\usepackage{array}
\usepackage{multirow}
\usepackage{hyperref}
\usepackage{verbatim}

\usepackage[mediumspace,mediumqspace,squaren]{SIunits}

\graphicspath{{fig/}}

\originalTeX
\begin{document}

\title{Calculations of the cell survival rate
after irradiating with minibeams of
protons and $^{12}$C}
\maketitle
\authors{A.O.\,Svetlichnyi$^{a,}$\footnote{E-mail: svetlichnyi@inr.ru}, 
S.D.\,Savenkov$^{a,}$\footnote{E-mail: savenkov.sd@phystech.edu}
I.A.\,Pshenichnov$^{a,}$\footnote{E-mail: pshenich@inr.ru}}
\setcounter{footnote}{0}
\from{$^{a}$\,Institute for Nuclear Research of the Russian Academy of Sciences, 7a~Prospekt 60-letiya Oktyabrya, 117312, Moscow, Russia }

\begin{abstract}

The propagation of minibeams of protons and $^{12}$C in a water phantom was modelled with Geant4 v10.3, and the survival probabilities of human salivary gland cells representing healthy and tumour tissues of normal radiosensitivity were calculated with the modified microdosimetric kinetic model. The advantage of minibeams over homogeneous irradiation in sparing healthy tissues proximal to the tumour was demonstrated. Survival-volume histograms were proposed to quantify the difference in the survival probability of cells inside and outside the minibeam spots.   
\end{abstract}
\vspace*{6pt}


\label{sec:intro}
\section*{Introduction}

Proton and carbon-ion therapies are very effective in treating deep-seated solid tumours~\cite{Graeff2023}, but some collateral damage to the healthy tissues proximal to the target tumour volume is inevitable. Proton minibeam radiation therapy (pMBRT) with spatial fractionation of the dose field at the entrance to the patient's body is considered promising for mitigating such complications~\cite{Mazal2020}. In this method a homogeneous dose field in healthy tissues is replaced by arrays of parallel sub-millimetre-wide (0.3--0.7~mm FWHM) minibeams~\cite{Lamirault2020} with 1--2~mm distance between the minibeam centres. The lateral dose distribution of each individual proton minibeam gradually widens with depth due to multiple Coulomb scattering and secondary particle production in the tissues. Finally, the dose fields of the proton minibeams overlap in the target volume, providing effective tumour control with a solid homogeneous field. In contrast, the limited broadening of carbon-ion minibeams does not provide a sufficiently homogeneous dose field in the target volume located at the same depth. This motivated the use of two (four) crossed arrays of planar carbon minibeams from orthogonal directions to achieve a homogeneous dose field as these minibeams interleave in the target volume~\cite{Dilmanian2015a}.

In evaluating the efficiency of carbon-ion minibeam therapy with respect to proton minibeam therapy, it is necessary to consider the higher relative biological efficiency (RBE) of $^{12}$C compared to protons. In this work, the survival probability of human salivary gland (HSG) cells, representing both healthy tissues and tumour of normal radiosensitivity, irradiated with arrays of circular minibeams of protons and $^{12}$C was calculated on the basis of Monte-Carlo modelling with Geant4~\cite{Allison2016} and the modified microdosimetric kinetic (MK) model~\cite{Kase2006}. 

\section{Simulations of minibeam propagation in water and cell survival}\label{sec:model}

The propagation of protons and $^{12}$C in a $10\times 10\times 200$~mm$^3$ water phantom was simulated by means of the Geant4 toolkit~\cite{Allison2016} of version 10.3. All electromagnetic processes were modelled with the G4EMStandard\verb|_|opt3 physics list, while the Binary Cascade (G4BIC) and Quantum Molecular Dynamics (G4QMD) models were used to describe, respectively,  nuclear reactions induced by protons (neutrons) and hadronic nucleus-nucleus collisions. This approach has been validated previously~\cite{Pshenichnov2024} by comparing the calculated depth and lateral distributions of dose from pencil-like beams of protons and $^{12}$C in water with measured dose profiles. 

In this work the arrays of 16 parallel circular minibeams of protons and $^{12}$C of 0.5~mm FWHM at the entrance to the water phantom were modelled. For consistency with our previous study~\cite{Pshenichnov2024} focused on the calculations of the dose-volume histograms at the entrance and in the target volume, the minibeams were placed on rectangular and hexagonal grids with a centre-to-centre distance of 2~mm.  As in Refs.~\cite{Burigo2015,Pshenichnov2024}, sets of beam energies were adopted in simulations to obtain 60~mm wide spread-out Bragg peaks (SOBP) centred at 130~mm both for protons and $^{12}$C to deliver $5.1$~Gy$\times$RBE to the target volume approximately corresponding to 10\% of survival of tumour cells. Calculations were made also for homogeneous transverse dose distributions presently used in therapy with protons and $^{12}$C. Typically, up to 14 and 3 million histories of incident protons and $^{12}$C were simulated, respectively, to obtain three-dimensional distributions of dose and saturation-corrected dose-mean lineal energy $y^*$ in the phantom. The values of $y^*$ were calculated in each $0.1\times 0.1\times 0.1$~mm$^3$ voxel by integrating the probability density function $f(y)$ of the lineal energy $y$ of primary and secondary particles representing mixed radiation field:
\begin{equation}
y^* = \frac{ y_0^2 \int_{0}^{\infty} \left( 1-\exp{\left(-y^2/y_0^2\right)}\right) f(y){\rm d}y }{ \int_{0}^{\infty} yf(y) {\rm d} y } \ ,
\label{eq:ystar}
\end{equation}
with the parameter $y_0=150$~keV/{\micro\meter}~\cite{Kase2006} accounting for the saturation of biological effects induced by particles with very high lineal energy transfer (LET). 

The spatial distributions of cell survival $S$  were calculated by means of the  linear quadratic model (LQM) as $S = \exp(-\alpha D - \beta D^2)$.  According to the modified MK model~\cite{Kase2006}, the parameter $\alpha$ was calculated from $y^*$ given by Eq.~(\ref{eq:ystar}) as:   
\begin{equation}
\alpha = \alpha_0 + \frac{\beta}{\rho \pi r_d^2} y^* \ .
\label{eq:alpha}
\end{equation}
The following model parameters were taken for HSG cells~\cite{Kase2006}: $\alpha_0$ = 0.13~Gy$^{-1}$ representing the initial slope of the survival fraction curve in the limit of zero LET, $\beta = 0.05$~Gy$^{-2}$, $\rho = $1~g/cm$^3$ as the density of tissue and $r_d = $~\unit{0.42}{\micro\meter} as the radius of a sub-cellular domain in the modified MKM. 
In this approach, see Ref.~\cite{Burigo2015} for details, the parameters $\alpha$ and $\beta$ were calculated individually in each phantom voxel to characterize highly non-homogeneous mixed radiation field of minibeams. 

\section{Cell survival probability after the irradiation with minibeams of protons and $^{12}$C}

\begin{figure}[htb!]
    \centering
    \begin{minipage}{0.49\linewidth}
    \includegraphics[width=1.0\linewidth]{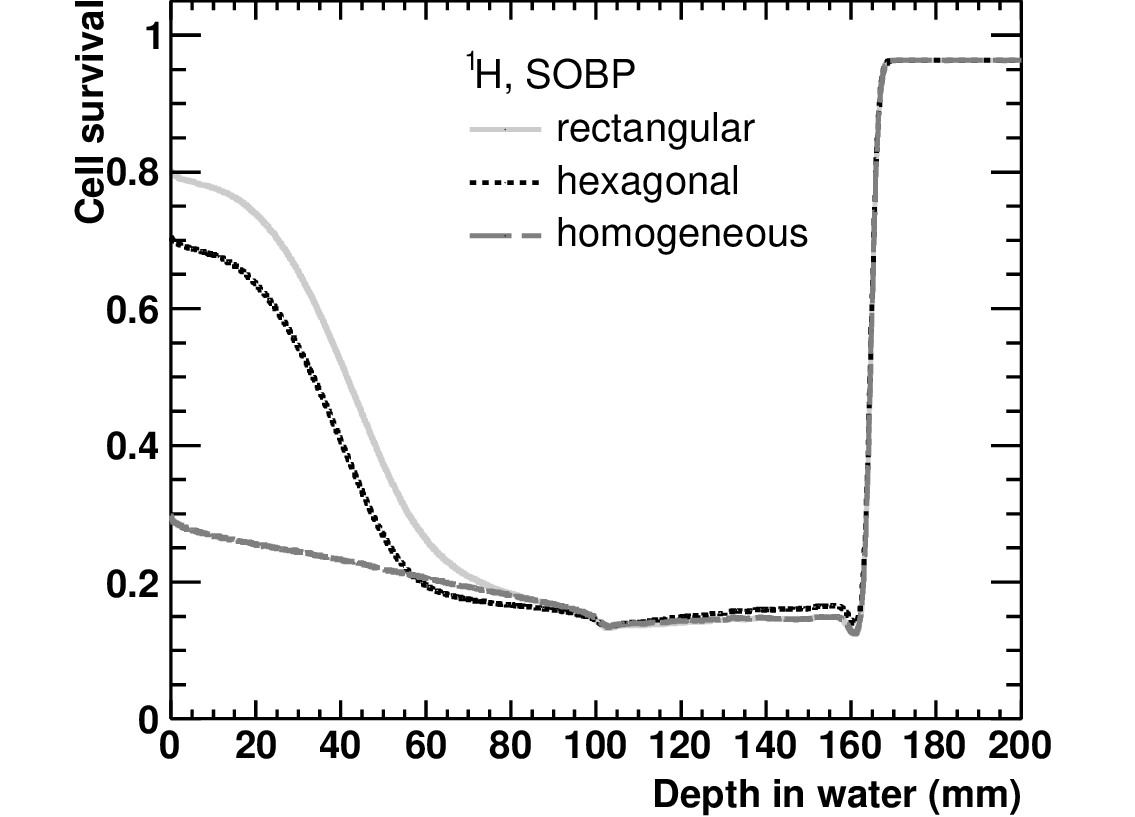}
    \end{minipage}
    \begin{minipage}{0.49\linewidth}
    \includegraphics[width=1.0\linewidth]{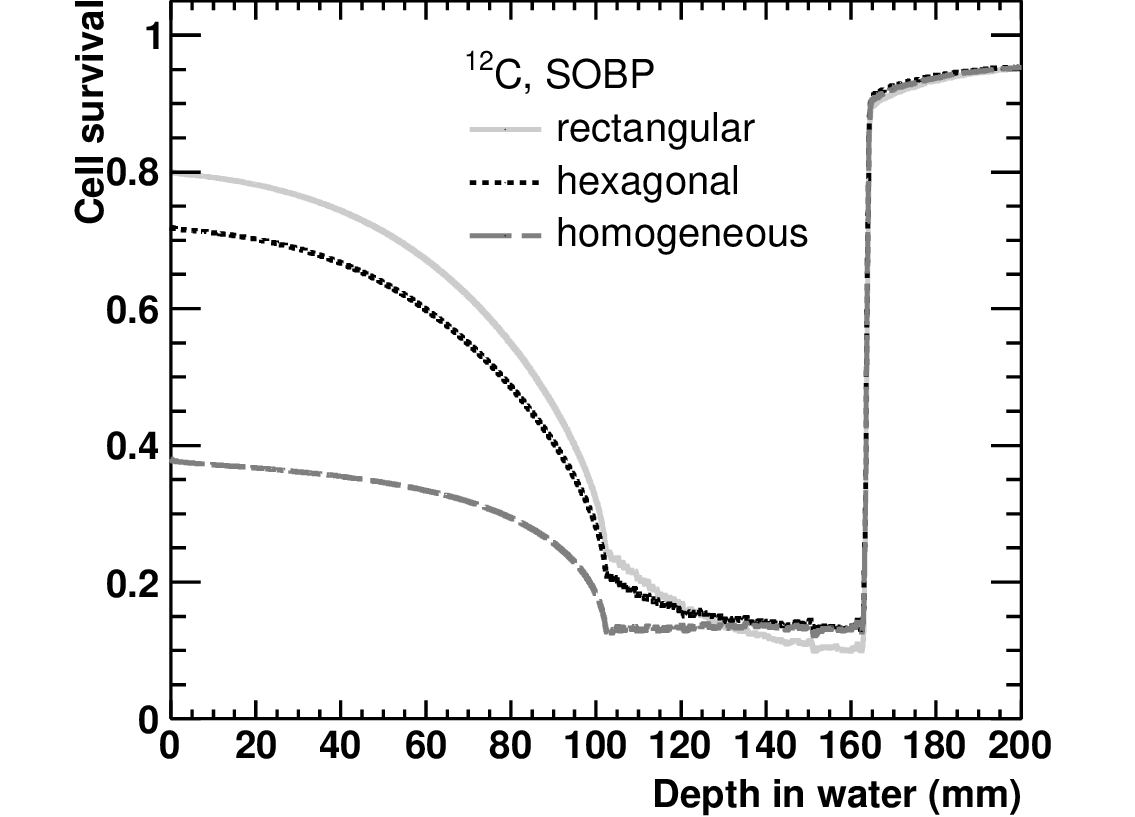}
    \end{minipage}
    \caption{Average survival probability as a function of depth in phantom after the irradiation of HSG cells by protons (left) and $^{12}$C (right). The results for arrays of circular minibeams of 0.5~mm FWHM arranged on rectangular and hexagonal grids are shown by solid light and dotted dark lines, respectively. The results for homogeneous dose fields are shown by light dashed lines.}
    \label{fig:depth_survival}
\end{figure}

The average survival probability of HSG cells calculated in the central region of $2.6\times2.6$~mm$^2$ of the transverse plane as a function of depth in the phantom 
is shown in Fig.~\ref{fig:depth_survival}.
In this figure, the results for protons and $^{12}$C are presented, and the advantages of the minibeams compared to a homogeneous dose field can be seen. About 60--80\% of the cells survive at the entry of the phantom (0--20~mm depth) after irradiation with 0.5~mm FWHM minibeams arranged on rectangular and hexagonal grids. In contrast, the probability of survival is twice as low (30--40\%) in the same depth interval for the homogeneous field. Nevertheless, it was calculated that approximately the same large fraction of cells (85--90\%) is inactivated by the minibeams and uniform field in the SOBP (100--160~mm depth) corresponding to the target tumour volume.

\begin{figure}[htb!]
    \centering
    \includegraphics[width=1.0\linewidth]{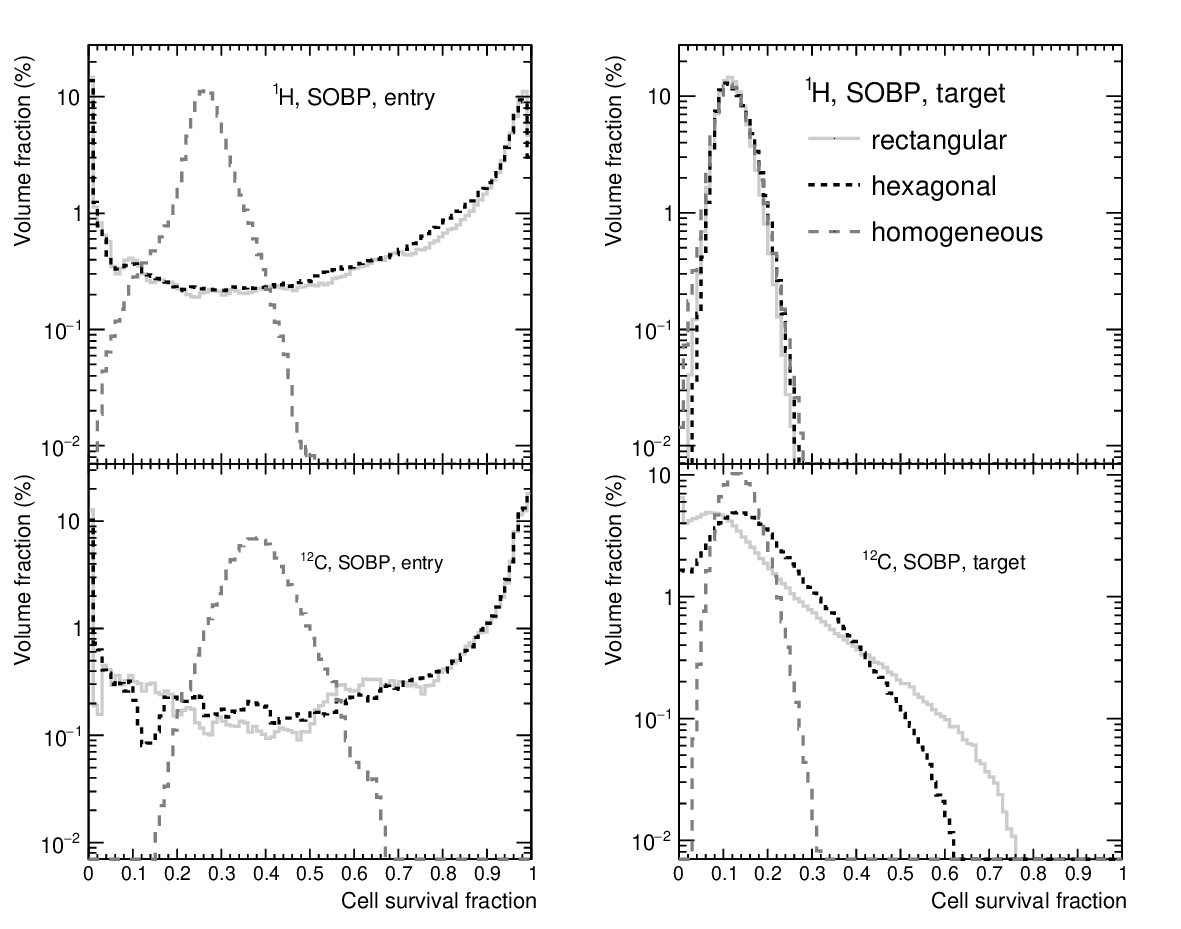}
    \caption{Fractions of the entry (left) and target (right) volumes characterized by a given cell survival fraction after the irradiation with proton (top) and $^{12}$C (bottom) minibeams and homogeneous dose fields. The notations are the same as in Fig.~\ref{fig:depth_survival}.}
    \label{fig:svh}
\end{figure}

The sparing effect of the minibeams at the entrance is associated with the dose-volume effect~\cite{Dilmanian2015a}. This motivates the calculation of differential survival volume histograms (SVHs) for proton and $^{12}$C irradiation shown in Fig.~\ref{fig:svh}. The SVHs represent the fractions of the entry and target volumes characterised by a given survival fraction $S$. The SVHs calculated in the entry volumes are bimodal and almost identical for both minibeam irradiation geometries for protons as well as for $^{12}$C. The cells are inactivated only in $\sim 10$\% of the entry volumes because they are directly covered by the minibeams. However, the cells are not affected at all in other 10\% of the volumes, and more than 50\% of the cells survive elsewhere in the entry volume. In contrast, after homogeneous irradiation there are no parts of the entrance volumes with $S>0.7$. In the target volume, all three SVHs for protons coincide because the converging proton minibeams provide a homogeneous dose field there.

In contrast to the proton minibeams, the $^{12}$C minibeams do not completely overlap in the target volume, especially at the proximal edge of the SOBP, and the degree of overlap is different for rectangular and hexagonal grids. Therefore, the target SVHs for two minibeam geometries do not coincide, and they are also very different from the SVH in the homogeneous field with $S<0.3$, despite the similarity of the average $S$ shown in Fig.~\ref{fig:depth_survival}.  As can be seen in Fig.~\ref{fig:svh}, the domains with $S>0.3$ remain after irradiation with minibeams due to their incomplete overlap. 

\section*{Conclusion}\label{sec:conclusion}

The much higher average survival probability calculated for HSG cells at the entrance to the phantom demonstrates the advantages of proton and $^{12}$C minibeams over homogeneous irradiation. This conclusion is supported by the first calculation of differential survival volume histograms (SVHs) in the context of external beam radiotherapy. The bimodal shapes of the SVHs at the entrance to the phantom quantify the difference in the survival probability for cells inside and outside the minibeam spots.

\section*{Acknowledgments}
This study was funded by the Russian Science Foundation (RSF) grant No. 23-25-00285 ”Modeling of the physical and biological properties of therapeutic minibeams of protons and nuclei”. The authors are grateful to the RSF for the support.

\bibliographystyle{pepan}
\bibliography{Svetlichnyi_minibeams_survival}

\end{document}